\documentclass[twocolumn,aps,showpacs,floatfix,superscriptaddress]{revtex4}
\usepackage{amsmath,amssymb,eucal,graphicx}
\begin{document}

\title{On the Role of Global Warming on the Statistics of
  Record-Breaking Temperatures}
\author{S. Redner}\email{redner@bu.edu}
\altaffiliation{Most of this work was completed when the first author was on
  leave at the Center for Nonlinear Studies and Theoretical Division, Los
  Alamos National Laboratory, Los Alamos, New Mexico 87545, USA}
\affiliation{Center for Polymer Studies \&
Department of Physics, Boston University, Boston, Massachusetts, 02215 USA}
\author{Mark R.~Petersen}\email{mpetersen@lanl.gov}
\affiliation{Computer and Computational Sciences Division 
and Center for Nonlinear Studies,
Los Alamos National Laboratory, Los Alamos, New Mexico, 87545 USA}

\begin{abstract}
  We theoretically study the statistics of record-breaking daily temperatures
  and validate these predictions using both Monte Carlo simulations and 126
  years of available data from the city of Philadelphia.  Using extreme
  statistics, we derive the number and the magnitude of record temperature
  events, based on the observed Gaussian daily temperature distribution in
  Philadelphia, as a function of the number of years of observation.  We then
  consider the case of global warming, where the mean temperature
  systematically increases with time.  Over the 126-year time range of
  observations, we argue that the current warming rate is insufficient to
  measurably influence the frequency of record temperature events, a
  conclusion that is supported by numerical simulations and by the
  Philadelphia data.  We also study the role of correlations between
  temperatures on successive days and find that they do not affect the
  frequency or magnitude of record temperature events.
\end{abstract}
\pacs{92.60.Ry, 92.60.Wc, 92.70.-j, 02.50.Cw} 
\maketitle

\section{INTRODUCTION}

Almost every summer, there is a heat wave somewhere in the US that garners
popular media attention \cite{Roach05ngn}.  During such hot spells, daily
record high temperatures for various cities are routinely reported in local
news reports.  A natural question arises: is global warming the cause of such
heat waves or are they merely statistical fluctuations?  Intuitively,
record-breaking temperature events should become less frequent with time if
the average temperature is stationary.  Thus it is natural to be concerned
that global warming is playing a role when there is a proliferation of
record-breaking temperature events.  In this work, we investigate how
systematic climatic changes, such as global warming, affect the magnitude and
frequency of record-breaking temperatures.  We then assess the potential role
of global warming by comparing our predictions both to record temperature
data and to Monte Carlo simulation results.

It bears emphasizing that record-breaking temperatures are distinct from
threshold events, defined as observations that fall outside a specified
threshold of the climatological temperature distribution \cite{Yan_ea02cc}.
Thus, for example, if a city's record temperature for a particular day is
$40^\circ$C, then an increase in the frequency of daily temperatures above
$36^\circ$C ({\it i.e.}, above the $90^{\rm th}$ percentile) is a threshold
event, but not a record-breaking event.  Trends in threshold temperature
events are also impacted by climate change, and is thus an active research
area
\cite{Yan_ea02cc,Mearns_ea84jcam,Hansen_ea88jgr,Katz_Brown92cc,Columbo_ea99jc,Unkasevic_ea05tac}.
Studying threshold events is also one of the ways to assess agricultural,
ecological, and human health effects due to climate change
\cite{Meehl_ea00bams,IPCC01bk}.

Here we examine the complementary issue of record-breaking temperatures, in
part because they are popularized by the media during heat waves and they
influence public perception of climate change, and in part because of the
fundamental issues associated with record statistics.  We focus on daily
temperature extremes in the city of Philadelphia, for which data are readily
available on the Internet for the period 1874--1999 \cite{PSU}.  In
particular, we study how temperature records evolve in time for each {\em
  fixed\/} day of the year.  That is, if a record temperature occurs on
January 1, 1875, how long until the next record on January 1 occurs?  Using
the fact that the daily temperature distribution is well approximated by a
Gaussian (Sec.~\ref{T-data}), we will apply basic ideas from extreme value
statistics in Sec.~\ref{ETR} to predict the magnitude of the temperature jump
when a new record is set, as well as the time between successive records on a
given day.  These predictions are derived for an arbitrary daily temperature
distribution, and then we work out specific results for the idealized case of
an exponential daily temperature distribution and for the more realistic
Gaussian distribution.

Although individual record temperature events are fluctuating quantities, the
average size of the temperature jumps between successive records and the
frequency of these records are systematic functions of time (see, {\it e.g.},
\cite{vonStorch_Zwiers99bk} for a general discussion).  This systematic
behavior permits us to make meaningful comparisons between our theoretical
predictions, numerical simulations (Sec.~\ref{simulations}), and the data for
record temperature events in Philadelphia (Sec.~\ref{TRD}).  Clearly, it
would be desirable to study long-term temperature data from many locations to
discriminate between the expected number of record events for a stationary
climate and for global warming.  For U.S. cities, however, daily temperature
records extend back only 100--140 years
\cite{Reitan_Moran77mwr,Balling_ea90jc}, and there are both gaps in the data
and questions about systematic effects caused by ``heat islands'' for
observation points in urban areas.  In spite of these practical limitations,
the Philadelphia data provide a useful testing ground for our theoretical
predictions.

In Sec.~\ref{SCT}, we investigate the effect of a slow linear global warming
trend \cite{ABN,trend} on the statistics of record-high and record-low
temperature events.  We argue that the presently-available 126 years of data
in Philadelphia, coupled with the current global warming rate, are
insufficient to meaningfully alter the frequency of record temperature events
compared to predictions based on a stationary temperature.  This conclusion
is our main result.  Finally, we study the role of correlations in the daily
temperatures on the statistics of record temperature events in
Sec.~\ref{corr}.  Although there are substantial correlations between
temperatures on nearby days and record temperature events tend to occur in
streaks, these correlations do not affect the frequency of record temperature
events for a given day.  We summarize and offer some perspectives in
Sec.~\ref{disc}.

\section{TEMPERATURE OBSERVATIONS}

The temperature data for Philadelphia were obtained from a website of the
Earth and Mineral Sciences department at Pennsylvania State University \cite{PSU}.
The data contain both the low and high temperatures in Philadelphia for each
day between 1874 and 1999.  The data are reported as an integer in degrees
Fahrenheit, so we anticipate an error of $\pm 1^\circ$F.  No information is
provided about the accuracy of the measurement or the precise location where
the temperature is measured.  Thus there is no provision for correcting for
the heat island effect if the weather station is in an increasingly urbanized
location during the observation period.  For each day, we also document the
middle temperature, defined as the average of the daily high and daily low.

To get a feeling for the nature of the data, we first present basic
observations about the average annual temperature and the variation of the
temperature during a typical year.

\subsection{Annual averages and extremes}

\begin{figure}[ht]
  \vspace*{0.cm} \includegraphics*[width=0.4\textwidth]{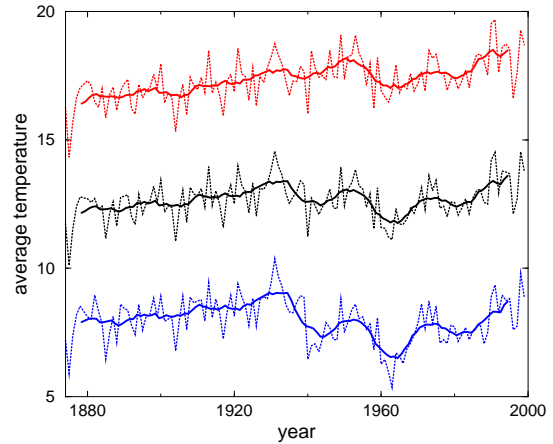}
  \caption{(Color online) Average annual high, middle, and low temperature
    (in degrees Celsius) for each year between 1874 and 1999 (jagged dotted
    lines).  Also shown are the corresponding 10-year averages (solid
    curves).
    \label{av-temps}}
\end{figure}

Figure~\ref{av-temps} shows the average annual high, middle, and low
temperature for each year between 1874 and 1999.  To help discern systematic
trends, we also plot 10-year averages for each data set.  The average high
temperature for each year is increasing from 1874 until approximately 1950
and again after 1965, but is decreasing from 1950 to 1965.  Over the 126 years
of data, a linear fit to the time dependence of the annual high temperature
for Philadelphia gives an increase of $1.62^\circ$C, compared to the
well-documented global warming rate of $0.6\pm 0.2^\circ$C over the past
century \cite{IPCC01bk}.  On the other hand, there does not appear to be a
systematic trend in the dependence of the annual low temperature on the year.
A linear fit to these data give a {\it decrease} of $-0.38^\circ$C.  This
disparity between high and low temperatures is a puzzling and as yet
unexplained feature of the data.

\begin{figure}[ht]
  \vspace*{0.cm} \includegraphics*[width=0.4\textwidth]{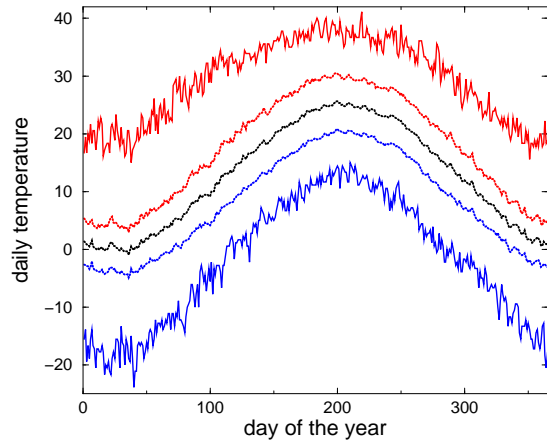}
  \caption{(Color online) Record high, average high, middle, and low, and
    record low temperature (in degrees Celsius) for each day of the year.
    \label{daily-avs+recs}}
\end{figure}

A basic feature about the daily temperature is its approximately sinusoidal
annual variation (Fig.~\ref{daily-avs+recs}).  The coldest time of the year
is early February while the warmest is late July.  An amusing curiosity is
the discernible small peak during the period January 20--25.  This anomaly is
the traditional ``January thaw'' in the northeastern US where sometimes
snowpack can melt and a spring-like aura occurs before winter returns (see
\cite{Jan} for a detailed discussion of this phenomenon).

Also shown, in Fig.~\ref{daily-avs+recs}, are the temperature extremes for
each day.  The highest recorded temperature in Philadelphia of $41.1^\circ$C
($106^\circ$F) occurred on August 7, 1918, while the lowest temperature of
$-23.9^\circ$C ($-11^\circ$F) occurred on February 9, 1934.  Record
temperatures also fluctuate more strongly than the mean temperature because
there are only 126 years of temperature data.  As a result of this short
time span, some days of the year have experienced very few records and the
resulting current extreme temperature can be far from the value that is
expected on statistical grounds (see Sec.~\ref{TRD}).

\subsection{Daily  temperature distribution}
\label{T-data}

To understand the magnitude and frequency of daily record temperatures, we
need the underlying temperature distribution for each day of the year.
Because temperatures have been recorded for only 126 years, the temperature
distribution for each individual day is not smooth.  To mitigate this
problem, we aggregate the temperatures over a 9-day range and then use these
aggregated data to define the temperature distribution for the middle day in
this range.  Thus, for example, for the temperature distribution on January
5, we aggregate all 126 years of temperatures from January 1--9 (1134 data
points).  We also use the middle temperature for each day to define the
temperature distribution.

\begin{figure}[ht]
  \vspace*{0.cm} \includegraphics*[width=0.45\textwidth]{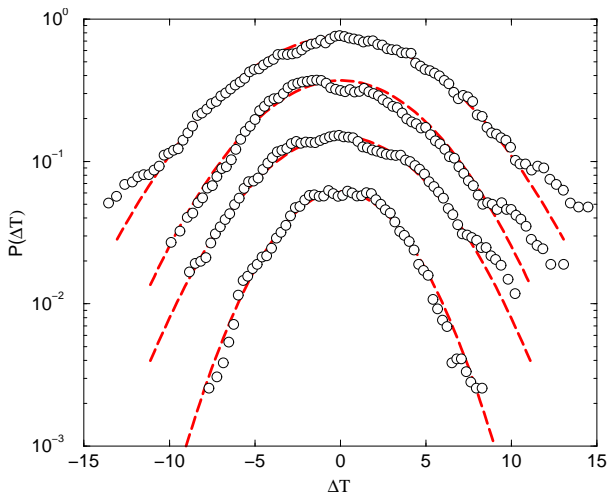}
  \caption{(Color online) Nine-day aggregated temperature distributions for
    January 5, April 5, October 5, and July 5 in degrees Celsius (top to
    bottom).  Each data set is averaged over a $10\%$ range---10, 9, 8, and 6
    points respectively, for January 5, April 5, October 5, and July 5.  The
    distributions are all shifted horizontally by the mean temperature for
    the day and then vertically to render all curves distinct.  The dashed
    curves are visually-determined Gaussian fits.  }
  \label{temp-dist-9day}
\end{figure}

Figure~\ref{temp-dist-9day} shows these aggregated temperature distributions
for four representative days---the $5^{\rm th}$ of January, April, July, and
October.  Each distribution is shifted vertically to make them all
non-overlapping.  We also subtracted the mean temperature from each of the
distributions, so that they are all centered about zero.  Visually, we obtain
good fits to these distributions with the Gaussian $P(\Delta T)\propto
e^{-(\Delta T)^2/2\sigma^2}$, where $\Delta T$ is the deviation of the
temperature from its mean value (in $^\circ$C), and with $\sigma\approx
5.07$, 4.32, 4.12, and 3.14 for January 5, April 5, October 5, and July 5,
respectively.  We therefore use a Gaussian daily temperature distribution as
the input to our investigation of the frequency of record temperatures in the
next section.

An important caveat needs to be made about the daily temperature
distribution.  Physically, this distribution cannot be Gaussian {\em ad
  infinitum}.  Instead, the distribution must cut off more sharply at finite
temperature values that reflect basic physical limitations (such as the
boiling points of water and nitrogen).  We will show in the next section that
such a cutoff strongly influences the average waiting time between successive
temperature records on a given day.

Notice that the width of the daily temperature distribution is largest in the
winter and smallest in the summer.  Another intriguing aspect of the daily
distributions is the tail behavior.  For January 5, there are deviations from
a Gaussian at both at the high- and low-temperature extremes, while for April
5 and October 5, there is an enhancement only on the high-temperature side.
This enhancement is especially pronounced on April 5, which corresponds to
the season where record high temperatures are most likely to occur (see
Sec.~\ref{disc} and Fig.~\ref{daily-var}).  What is not possible to determine
with 126 years of data is whether the true temperature distribution is
Gaussian up to the cutoff points and the enhancement results from relatively
few data, or whether the true temperature distribution on April 5 actually
has a slower than Gaussian high-temperature decay.

\section{EVOLUTION OF RECORD TEMPERATURES}
\label{ETR}

\begin{figure}[ht]
  \vspace*{0.cm}
  \includegraphics*[width=0.48\textwidth]{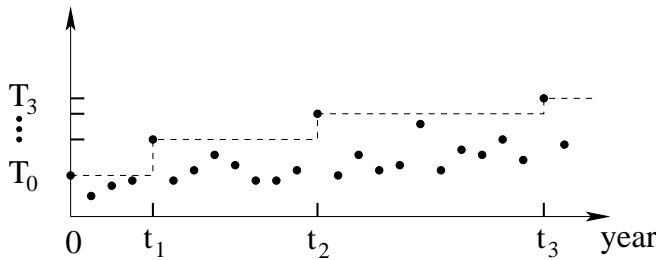} 
  \caption{Schematic evolution of the record high temperature on a specified
    day for each passing year.  Each dot represents the daily high temperature
    for different years.  The first temperature is, by definition, the zeroth
    record temperature $T_0$.  This event occurs in year $t_0=0$.  Successive
    record temperatures $T_1$, $T_2$, $T_3,\ldots$ occur in years $t_1$,
    $t_2$, $t_3,\ldots$.
    \label{record-evolution}}
\end{figure}

We now determine theoretically the frequency and magnitude of record
temperature events.  The schematic evolution of these two characteristics is
sketched in Fig.~\ref{record-evolution} for the case of record high
temperatures.  Each time a record high for a {\em fixed\/} day of the year is
set, we document the year $t_i$ when the $i^{\rm th}$ record occurred and the
corresponding record high temperature $T_i$.  Under the (unrealistic)
assumptions that the temperatures for each day are independent and identical,
we now calculate the average values of $T_i$ and $t_i$ and their underlying
probability distributions (For a general discussion of record statistics for
excursions past a fixed threshold, see {\it e.g.}  \cite{ABN,Vanm}, while
related work on the evolution of records is given in
Ref.~\cite{Schmittmann_Zia99ajp}).

Suppose that the daily temperature distribution is $p(T)$.  Two subsidiary
distributions needed for record statistics are: (i) the probability that a
randomly-drawn temperature {\em exceeds} $T$, $p_>(T)$, and (ii) the
probability that that this randomly-selected temperature {\em is less than}
$T$, $p_<(T)$.  These distributions are \cite{Galambos87bk}:
\begin{eqnarray}
\label{pg}
p_<(T)\equiv \int_0^T p(T')\, dT'\,; \quad
p_>(T)\equiv \int_T^\infty p(T')\, dT'.
\end{eqnarray}

We now determine the $k^{\rm th}$ record temperature $T_k$ recursively.  We
use the terminology of record high temperatures, but the same formalism
applies for record lows.  Clearly $T_0$ coincides with the mean of the daily
temperature distribution, $T_0\equiv \int_0^\infty T\, p(T)\, dT$.  The next
record temperature is the mean value of that portion of the temperature
distribution that lies beyond $T_0$: that is,
\begin{eqnarray}
\label{T1}
 T_1 \equiv \frac{\int_{T_0}^\infty  T\, p(T)\, dT} {\int_{T_0}^\infty  p(T)\, dT}\,.
\end{eqnarray}
This formula actually contains a sleight of hand.  More properly, we should
average the above expression over the probability distribution for $T_0$ to
obtain the true average value of $T_1$, rather than merely using the typical
or the average value of $T_0$ in the lower limit of the integral.
Equation~(\ref{T1}) therefore does not give the true average value of $T_1$, but
rather gives what we term the {\em typical\/} value of $T_1$.  We will show
how to compute the average value shortly.

Proceeding recursively, the relation between successive typical record
temperatures is given by
\begin{eqnarray}
\label{Tk-gen}
 T_{k+1} \equiv \frac{\int_{ T_{k}}^\infty  T\, p(T)\, dT}
{\int_{ T_{k}}^\infty  p(T)\, dT}\,,
\end{eqnarray}
where the above caveat about using the typical value of $T_k$ in the lower
limit, rather than the average over the (as yet) unknown distribution of
$T_k$, still applies.

We now compute $\mathcal{P}_k(T)$, the probability that the $k^{\rm th}$
record temperature equals $T$; this distribution is subject to the initial
condition $\mathcal{P}_0(T)=p(T)$.  For the $k^{\rm th}$ record temperature,
the following conditions must be satisfied (refer to
Fig.~\ref{record-evolution}): (i) the previous record temperature $T'$ must
be less than $T$, (ii) the next $n$ temperatures, with $n$ arbitrary, must
all be less than $T'$, and (iii) the last temperature must equal $T$.
Writing the appropriate probabilities for each of these events, we obtain
\begin{eqnarray}
\label{prob-T}
\mathcal{P}_k(T)&=& \left(\int_0^{T} \mathcal{P}_{k-1}(T')\sum_{n=0}^\infty 
[\,p_<(T')\,]^n \, dT'\right) p(T)
 \nonumber \\
&=& \left(\int_0^{T} \frac{\mathcal{P}_{k-1}(T')}{p_>(T')}
\, dT'\right) p(T)\,.
\end{eqnarray}
The above formula recursively gives the probability distribution for each record
temperature in terms of the distribution for the previous record.

Complementary to the magnitude of record temperatures, we determine the time
between successive records.  Suppose that the current record temperature
equals $T_k$ and let $q_n(T_k)$ be the probability that a new record
high---the $(k+1)^{\rm st}$---is set $n$ years later.  For this new record,
the first $n-1$ highs after the current record must all be less than $T_k$,
while the $n^{\rm th}$ high temperature must exceed $T_k$.  Thus
\begin{equation}
\label{qn}
q_n(T_k)= p_<(T_k)^{n-1}\, p_>(T_k)\,.
\end{equation}
The number of years between the $k^{\rm th}$ record high $T_k$ and the
$(k+1)^{\rm st}$ record $T_{k+1}$ is therefore
\begin{eqnarray}
\label{nav}
t_{k+1}-t_k =\sum_{n=1}^\infty n \,p_<^{n-1}\,\, p_>  = \frac{1}{p_>(T_k)}\,.
\end{eqnarray}
We emphasize that this waiting time gives the time between the $k^{\rm th}$
record and the $(k+1)^{\rm st}$ record when the $k^{\rm th}$ record
temperature equals the specified value $T_k$.  If the typical value of $T_k$
is used in Eq.~(\ref{nav}), we thus obtain a quantity that we term the
typical value of $t_k$.  

To obtain the true average waiting time, we first define $Q_n(k)$ as the
probability that the $k^{\rm th}$ record is broken after $n$ additional
temperature observations, averaged over the distribution for $T_k$.  Using
the definition of $q_n$, we obtain the formal expression
\begin{eqnarray}
\label{Qnk}
Q_n(k)&\equiv& \int_0^\infty \mathcal{P}_k(T)\,q_n(T) \, dT\nonumber \\
&=&\int_0^\infty \mathcal{P}_k(T)\, p_<(T)^{n-1}\, p_>(T)\, dT \,.
\end{eqnarray}
Different approaches to determine the $Q_n$ are given in
Refs.~\cite{ABN,Glick78amm}.

There are a number of fundamental results available about record statistics
that are {\em universal\/} and do not depend on the form of the initial daily
temperature distribution, as long as the daily temperatures are independent
and identically-distributed (iid) continuous variables
\cite{ABN,Glick78amm,Sibani,KJ,Benestad03cr}.  In a string of $n+1$
observations (starting at time $n=0$), there are $n!$ permutations of the
temperatures out of $(n+1)!$ total possibilities in which the largest
temperature is the last of the string.  Thus the probability that a new
record occurs in the $n^{\rm th}$ year of observation, $R_n$, is simply
\cite{ABN,Glick78amm,Sibani,KJ,Benestad03cr}
\begin{equation}
\label{Rn}
R_n=\frac{1}{n+1}.
\end{equation}

In a similar vein, the probability that the initial ($0^{\rm th}$) record is
broken at the $n^{\rm th}$ observation, $Q_n(0)$, requires that the last
temperature is the largest while the $0^{\rm th}$ temperature is the second
largest out of $n+1$ independent variables.  The probability for this event
is therefore
\begin{equation}
\label{Qn0-exact}
Q_n(0)=\frac{1}{n(n+1)}\,,
\end{equation}
again independent of the form of the daily temperature distribution.  Thus
the average waiting time between the zeroth and first records, $\langle
n\rangle=\sum_{n=1}^\infty n\, Q_n(0)$ is infinite!  

More generally, the distribution of times between successive records can be
obtained by simple reasoning \cite{Sibani,KJ}.  Consider a string of iid
random variables that are labeled by the time index $n$, with
$n=0,1,2,\ldots,t$.  Define the indicator function
\begin{equation}
\label{indi}
\sigma_n = 
\begin{cases}
1 &\text{\rm if record occurs in $n^{\rm th}$ year} \\
0 &\text{otherwise}.
\end{cases}
\end{equation}
By definition, the probability for a record to occur in the $n^{\rm th}$ year
is $R_n=\langle \sigma_n\rangle = \frac{1}{n+1}$.  Therefore the average
number of records that have occurred up to time $t$ is
\begin{equation}
\label{Rn-av}
\langle R_n\rangle = \sum_{n=1}^t\langle \sigma_n\rangle \sim \ln t.
\end{equation}
Moreover, because the order of all non-record events is immaterial in the
probability for a record event, there are no correlations between the times
of two successive record events.  That is, $\langle\sigma_m\sigma_n\rangle =
\langle\sigma_m\rangle\langle\sigma_n\rangle$.  Thus the probability
distribution of records is described by a Poisson process in which the mean
number of records up to time $t$ is $\ln t$.  Consequently, the probability
$\Pi(n,t)$ that $n$ records have occurred up to time $t$ is given by
\cite{Sibani}
\begin{equation}
\label{PRn}
\Pi(n,t) \sim \frac{(\ln t)^n}{n!}\, e^{-\ln t} = \frac{(\ln t)^n}{n!}\,\frac{1}{t}\, .
\end{equation}

To appreciate the implications of these formulae for record statistics, we
first consider the warm-up exercise of an exponential daily temperature
distribution.  For this case, all calculations can be performed explicitly
and the results provide intuition into the nature of record temperature
statistics.  We then turn to the more realistic case of the Gaussian
temperature distribution.

\subsection{Exponential distribution}

Suppose that the temperature distribution for each day of the year is
$p(T)=\mathcal{T}^{-1}\,e^{-T/\mathcal{T}}$.  Equation~(\ref{pg}) then gives
\begin{eqnarray}
\label{pg-exp}
p_<(T)= 1-e^{-T/\mathcal{T}}\,;\qquad
p_>(T)= e^{-T/\mathcal{T}}\,.
\end{eqnarray}
We now determine the typical value of each $T_k$.  The zeroth record
temperature is $T_0=\int_0^\infty T\, p(T)\, dT =\mathcal{T}$.  Performing
the integrals in Eq.~(\ref{Tk-gen}) successively for each $k$ gives the basic
result
\begin{equation}
\label{Tk}
 T_k = (k+1)\mathcal{T}\,,
\end{equation}
namely, a constant jump between typical values of successive record
temperatures.

For the probability distribution for each record temperature, we compute
$\mathcal{P}_k(T)$ one at a time for $k=0,1,2,\ldots$ using
Eq.~(\ref{prob-T}).  This gives the gamma distribution \cite{gamma}
\begin{eqnarray}
\label{prob-T-exp}
\mathcal{P}_k(T)= \frac{1}{k!}\frac{T^k}{\mathcal{T}^{k+1}} \, e^{-T/\mathcal{T}}\,.
\end{eqnarray}
This distribution reproduces the typical values of successive temperature
records given by Eq.~(\ref{Tk}); thus the typical and true average values for
each record temperature happen to be identical for an exponential temperature
distribution. The standard deviation of $\mathcal{P}_k(T)$ is given by
$\sqrt{\langle T^2\rangle - \langle T\rangle^2}= \mathcal{T}\sqrt{k+1}$, so
that successive record temperatures become less sharply localized as $k$
increases.

For the typical time between the $k^{\rm th}$ and $(k+1)^{\rm st}$ records,
Eq.~(\ref{nav}) gives
\begin{eqnarray}
\label{nav-exp}
t_{k+1}-t_k = \frac{1}{p_>(T_k)} = e^{T_k/\mathcal{T}}\,.
\end{eqnarray}
Substituting $T_k=(k+1)\mathcal{T}$ into Eq.~(\ref{nav-exp}), the typical
time is $e^{T_{k}/\mathcal{T}}= e^{(k+1)}$.  Thus records become less likely
as the years elapse.  Notice that the time between records does not depend on
$\mathcal{T}$ because of a cancellation between the size of the temperature
``barrier'' (the current record) and the size of the jump to surmount the
record.

For the distribution of waiting times between records, we first consider the
time between $T_0$ and $T_1$ in detail to illustrate our approach.
Substituting Eqs.~(\ref{pg-exp}) and (\ref{prob-T-exp}) into Eq.~(\ref{Qnk}),
this distribution is
\begin{eqnarray}
\label{Qn0-exp}
Q_n(0)=\frac{1}{\mathcal{T}}\int_0^\infty \! e^{-T/\mathcal{T}} (1\!-\!e^{-T/\mathcal{T}})^{n-1} 
e^{-T/\mathcal{T}}\, dT \,.
\end{eqnarray}
Performing this integral by parts gives the result of Eq.~(\ref{Qn0-exact}),
$Q_n(0)=1/[n(n+1)]$.

For later applications, however, we determine the large-$n$ behavior of
$Q_n(0)$ by an asymptotic analysis.  Defining $x=T/\mathcal{T}$, we rewrite
Eq.~(\ref{Qn0-exp}) for large $n$ as
\begin{eqnarray}
\label{Qn0-asymp}
Q_n(0)&=&\int_0^\infty  e^{-x} (1-e^{-x})^{n-1} \, e^{-x}\, dx\nonumber \\
 &\sim & \int_0^\infty e^{-2x}\, e^{-ne^{-x}}\, dx.
\end{eqnarray}
The double exponential in the integrand changes suddenly from 0 to 1 when $n=
e^x$, or $x= \ln n$.  To estimate $Q_n(0)$, we may omit the double
exponential in the integrand and simply replace the lower limit of the
integral by $\ln n$.  This approach immediately leads to $Q_n(0)\sim n^{-2}$,
in agreement with the exact result.

In general, the average waiting time between the $k^{\rm th}$ and $(k+1)^{\rm
  st}$ record is, from Eq.~(\ref{Qnk}),
\begin{eqnarray}
\label{Qnk-exp}
Q_n(k)\!=\!\int_0^\infty \!\!\!\frac{1}{k!} \frac{T^k}{\mathcal{T}^{k+1}} e^{-T/\mathcal{T}} 
(1\!-\!e^{-T/\mathcal{T}})^{n\!-\!1} 
e^{-T/\mathcal{T}} dT.
\end{eqnarray}
While we can express this integral exactly in terms of derivatives of the
$\beta$ function \cite{Abramowitz_Stegun72bk}, it is more useful to determine
its asymptotic behavior by the same analysis as that given in
Eq.~(\ref{Qn0-asymp}).  We thus rewrite $(1-e^{-x})^{n-1}$ as a double
exponential and use the fact that this function is sharply cut off for $x<\ln
n$ to reduce the integral of Eq.~(\ref{Qnk-exp}) to
\begin{equation}
\label{Qnk-asymp}
Q_n(k)\sim\int_{\ln n}^\infty \frac{x^k}{k!} \,\,e^{-2x}\, dx\,.
\end{equation}
To find the asymptotic behavior of this integral, we note that the integrand
has a maximum at $x^*=k/2$.  Thus for $n>x^*$, the exponential decay term
controls the integral and we may again estimate its value by taking the
integrand at the lower limit to give $Q_n(k)\propto (\ln n)^k/n^2$.  As a
result of the power-law tail, the average waiting time between {\em any} two
consecutive records is infinite.

However, the observationally meaningful quantity is the typical value of the
waiting time and we thus focus on typical values to characterize the steps
between successive records depicted in Fig.~\ref{record-evolution}.  The
typical time to reach the $k^{\rm th}$ record, $t_k$, is simply the sum of
the typical times between records.  Thus
\begin{eqnarray}
\label{tk-sum}
t_k&=&(t_k-t_{k-1})+(t_{k-1}-t_{k-2}) + \ldots + (t_2-t_1)+t_1\nonumber\\ 
 &=& e^k+e^{k-1} + \ldots + e^2 + e^1\nonumber \\
&=&\frac{e^k-1}{1-e^{-1}} \approx 1.58 e^k\,.
\end{eqnarray}
Equivalently, $\ln\, t_k\approx k +0.459$ so that Eq.~(\ref{Tk}) gives
$T_k\approx (\ln t_k +0.541)\mathcal{T}.$ Therefore the $k^{\rm th}$ record
high temperature increases logarithmically with the total number of
observations, as expected from basic extreme statistics considerations
\cite{Galambos87bk}.

After $k$ record temperatures for a given day have been set, the probability
for the next record to occur is $p_>(T_k)= e^{-T_k/\mathcal{T}}$.  Since
$T_k\approx \mathcal{T}\,\ln t_k$, we recast this probability as a function
of time to obtain
\begin{equation}
  p_>(t)=e^{-T_k/\mathcal{T}}\propto e^{-\ln t}=1/t\,, 
\end{equation}
thus reproducing the general result in
\cite{ABN,Glick78amm,Sibani,KJ,Benestad03cr}.  The annual number of record
temperatures after $t$ years should be $365/t$; for the Philadelphia data,
this gives 2.90 record temperatures for the year 2000, 126 years after the
start of observations.

\subsection{Gaussian distribution}

We now study record temperature statistics for the more realistic case of a
Gaussian daily temperature distribution.  Again, to avoid the divergence
caused the unphysical infinite limits in the Gaussian, we begin by computing
the typical value $T_k$ of the $k^{\rm th}$ record temperature, and the
typical time $t_k$ until this record.  While the calculational steps to
obtain these quantities are identical to those of the previous subsection,
the details are more complicated because the integrals for $p_<$ and $p_>$
must be evaluated numerically or asymptotically.

As will become evident, the mean value in the Gaussian merely sets the value
of $T_0$ and plays no further role in successive record temperatures.  Thus
for the daily temperature distribution, we use the canonical form
\begin{equation}
\label{p-G}
  p(T)=\frac{1}{\sqrt{2\pi \sigma^2}}\, e^{-T^2/2\sigma^2}
\end{equation}
to determine the values of successive record temperatures.  The exceedance
probability then is
\begin{eqnarray}
\label{pg-G}
p_>(T) \!\!&=&\!\! \int_T^\infty \!\!\frac{1}{\sqrt{2\pi \sigma^2}}\, e^{-x^2/2\sigma^2}
dx\nonumber 
= \frac{1}{2}\,{\rm erfc}\Big(T/\sqrt{2\sigma^2}\Big) \nonumber \\
&\sim& \frac{1}{\sqrt{2\pi}}\,\frac{\sigma}{T}\,e^{-T^2/2\sigma^2}\,\quad
T\gg\sqrt{2\sigma^2}\,,
\end{eqnarray}
where erfc$(z)$ is the complementary error function
\cite{Abramowitz_Stegun72bk}.

Clearly, $T_0 =0$, since the Gaussian distribution is symmetric.  If we had
used a Gaussian with a non-zero mean value, then all the $T_k$ would merely
be shifted higher by this mean value.  For the next record temperature,
Eq.~(\ref{Tk-gen}) gives
\begin{equation}
\label{T1G}
 T_1 =
\frac{\int_0^\infty \frac{1}{\sqrt{2\pi \sigma^2}}\, T\, e^{-T^2/2\sigma^2}\,dT}
{\int_0^\infty \frac{1}{\sqrt{2\pi \sigma^2}}\, e^{-T^2/2\sigma^2}\,dT}\,.
\end{equation}
Substituting $u=T^2/2\sigma^2$ and $v=T/\sqrt{2\sigma^2}$ in the numerator
and denominator, respectively, we obtain
\begin{eqnarray}
\label{T1-result}
 T_1 =
\frac{\int_0^\infty \frac{\sigma}{\sqrt{2\pi}}\, e^{-u}\,du}
{\frac{1}{2}\,{\rm erfc}(0)}
= \sqrt{\frac{2}{\pi}}\, \sigma \,.
\end{eqnarray}
Continuing this recursive computation, Eq.~(\ref{Tk-gen}) gives
\begin{eqnarray}
\label{T-recur-G}
T_{k+1} &=& \frac{
\int_{ T_k}^\infty \frac{1}{\sqrt{2\pi \sigma^2}}\, T\, e^{-T^2/2\sigma^2}\,dT}
{\int_{\ T_k}^\infty \frac{1}{\sqrt{2\pi \sigma^2}}\, e^{-T^2/2\sigma^2}\,dT}
\nonumber \\
&=&\frac{  T_1\, e^{-T_k^2/2\sigma^2}} 
{{\rm erfc}\big(T_k/\sqrt{2\sigma^2}\big)}\,.
\end{eqnarray}

For the first few $k$, it is necessary to evaluate the error function
numerically and we find $ T_2\approx 1.712\, T_1$, $T_3\approx 2.288\, T_1$,
$T_4 = 2.782\, T_1$, {\it etc}.  Now from Eq.~\eqref{T1-result}, the argument
of the error function in Eq.~\eqref{T-recur-G} is $T_k/\sqrt{2\sigma^2}=
T_k/(T_1\sqrt{\pi})$.  Thus for $k\geq 3$, this argument is greater than 1,
and it becomes increasingly accurate to use the large-$z$ asymptotic form
\cite{Abramowitz_Stegun72bk}
\begin{eqnarray*}
{\rm erfc}(z)\sim \frac{e^{-z^2}}{z\sqrt{\pi}}\left(1-\frac{1}{2z^2} +\ldots\right)\,.
\end{eqnarray*}
This approximation reduces the recursion for $T_{k+1}$ to
\begin{eqnarray}
\label{Tk-asymp}
T_{k+1} &=& \frac{ T_1\, e^{-T_k^2/2\sigma^2}}
{{\rm erfc}\big(T_k/\sqrt{2\sigma^2}\big)}\nonumber \\
&\sim& \frac{  T_1\, e^{ -T_k^2/2\sigma^2}}
{\sqrt{\frac{2\sigma^2}{\pi T_k^2}}\, e^{-T_k^2/2\sigma^2}\,
\left(1-\frac{1}{2(T_k/\sqrt{2\sigma^2})^2}+\ldots\right)} \nonumber \\
&\sim& T_k\left(1+\frac{\sigma^2}{T_k^2}\right)\,,
\end{eqnarray}
where we have used $T_1=\sqrt{2\sigma^2/\pi}$ from Eq.~(\ref{T1-result}).

Writing the last line as $T_{k+1}-T_k= \sigma^2/T_k$, approximating the
difference by a derivative, and integrating, the $k^{\rm th}$ record
temperature for large $k$ has the remarkably simple form
\begin{equation}
\label{Tk-G}
T_k\sim \sqrt{2k\sigma^2}\,.
\end{equation}
Thus successive record temperatures asymptotically become more closely spaced
for the Gaussian distribution.  It should be noted, however, that the largest
number of record temperature events on any given day in the Philadelphia data is
10, so that the applicability of the asymptotic approximation is necessarily
limited.

The more fundamental measure of the temperature jumps is again
$\mathcal{P}_k(T)$, the probability distribution that the $k^{\rm th}$ record
high equals $T$.  For a Gaussian daily temperature distribution, the general
recursion given in Eq.~(\ref{prob-T}) for $\mathcal{P}_k(T)$ is no longer
exactly soluble, but we can give an approximate solution that we expect will
become more accurate as $k$ is increased.  We merely employ the large-$T$
asymptotic form for $p_>(T)$ in the recursion for $\mathcal{P}_k(T)$ even
when $k$ is small so that $T$ is not necessarily much larger than $\sigma$.
Using this approach, we thus obtain for $\mathcal{P}_1(T)$
\begin{eqnarray}
\label{P1-G}
\mathcal{P}_1(T)&\approx& \left(\int_0^T \frac{
\frac{1}{\sqrt{2\pi\sigma^2}} e^{-T'^2/2\sigma^2}}
{\sqrt{\frac{\sigma^2}{2\pi T'^2}} e^{-T'^2/2\sigma^2}}\, dT'  \right)
 \frac{1}{\sqrt{2\pi\sigma^2}}\, e^{-T^2/2\sigma^2} \nonumber \\
&\sim&\frac{T^2}{2\sigma^2} \, \frac{1}{\sqrt{2\pi\sigma^2}}\, e^{-T^2/2\sigma^2}\,.
\end{eqnarray}
Continuing this straightforward recursive procedure then gives
\begin{eqnarray}
\label{Pk-G}
\mathcal{P}_k(T)
\approx \frac{1}{\Gamma(k+\frac{1}{2})}\frac{T^{2k}}{\sigma^{2k+1}} \, 
e^{-T^2/2\sigma^2}\,,
\end{eqnarray}
where the amplitude is determined after the fact by demanding that the
distribution is normalized.

\begin{figure}[ht]
  \vspace*{0.cm}
  \includegraphics*[width=0.45\textwidth]{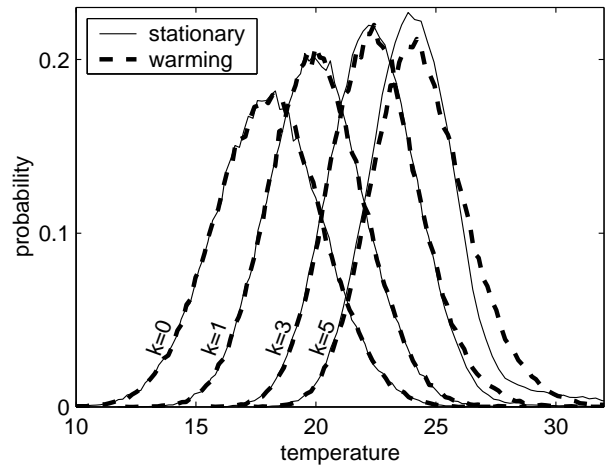}
  \caption{Simulation data for the probability distribution of the 
    $k^{\rm th}$ record high temperature in degrees Celsius, $\mathcal{P}_k(T)$.  The
    distribution $\mathcal{P}_0(T)$ coincides with the Gaussian of
    Eq.~(\ref{p-G}), whose parameters match the average temperature and
    dispersion in Philadelphia.  The solid curves correspond to a stationary
    temperature, while the dashed curves correspond to global warming with
    rate $v=0.012^\circ$C year$^{-1}$ (see Sec.~\ref{SCT}).
    \label{Pk}}
\end{figure}

In spite of the crudeness of this approximation, this distribution agrees
reasonably with our numerical simulation results shown in Fig.~\ref{Pk}
(details of the simulation are described in the following section).  The
distributions $\mathcal{P}_k(T)$ move systematically to higher temperatures
and become progressively narrower as $k$ increases, in accordance with naive
intuition.  The approximate form of Eq.~\eqref{Pk-G} gives a similar shape to
the simulated distributions, but there is an overall shift to higher
temperatures by roughly 1--2$^\circ$C.

Next, we study the typical time between successive record temperatures.
Equation~(\ref{nav}) states that $t_{k+1}-t_k =1/(p_>(T_k))$.  Using the above
asymptotic expansion of the complementary error function in the integral for
$p_>$ and $T_k\sim \sqrt{2k\sigma^2}$ from Eq.~(\ref{Tk-G}), we obtain, for
large $k$,
\begin{eqnarray}
\label{nav-G}
t_{k+1}-t_k
\sim \sqrt{4\pi}\frac{T_k}{\sqrt{2\sigma^2}} \, e^{T_k^2/2\sigma^2}
\sim  \sqrt{4\pi k}\, e^{k} \,.
\end{eqnarray}
Again, the times between records are independent of $\sigma$; this
independence arises because both the size of the record and the magnitude of
the jumps to surpass the record are proportional to $\sigma$, so that its
value cancels out in the waiting times.

Finally, we compute the asymptotic behavior for the distribution of waiting
times between records.  For simplicity, we consider only the waiting time
distribution $Q_n(0)$ until the first record.  The distribution of waiting
times for subsequent records has the same asymptotic tail as $Q_n(0)$, but
also contains more complicated pre-asymptotic factors.  Substituting the
Gaussian for $p(T)$ and the asymptotic form for $p_>(T)$ into
Eq.~(\ref{Qnk}), and then expanding $(1-p_>)^{n-1}$ as a double exponential,
we obtain
\begin{eqnarray}
\label{Qn0-G}
Q_n(0)\!\sim\!\!  \int_0^\infty\!\!\! \frac{1}{2\pi x}\,
\exp\!\!\left[-\frac{x^2}{\sigma^2}-\!\sqrt{\frac{n^2\sigma^2}{2\pi x^2}}
\,e^{-x^2\!/2\sigma^2}\right]\! dx.
\end{eqnarray}
The double exponential again cuts off the integral when $x$ is less than a
threshold value $x*\sim \sqrt{2\sigma^2\ln n}$.  As a result,
Eq.~(\ref{Qn0-G}) reduces to
\begin{eqnarray}
\label{Qn0-G-asymp}
Q_n(0)&\sim& \int_{\sqrt{2\sigma^2 \ln n}}^\infty\,\,\frac{1}{2\pi x}\,  
e^{-x^2/\sigma^2}\, dx \sim \frac{1}{n^{2}}\,.
\end{eqnarray}
In the final result, we drop logarithmic corrections because the
approximation made in writing Eq.~(\ref{Qn0-G}) also contains errors of the
same magnitude.  Thus the distribution of waiting times $n$ until the first
record again has a $n^{-2}$ power-law tail and the mean waiting time is
infinite.

The typical time until the $k^{\rm th}$ record is again given by the
sum of successive time intervals.  Asymptotically, Eq.~(\ref{nav-G}) gives
\begin{equation}
t_k\sim \int_0^k \sqrt{4\pi n}\,\, e^n\, dn \sim \sqrt{4\pi k}\,\, e^k\,,
\end{equation}
or $k\approx \ln t -\frac{1}{2}\ln(4\pi\ln t)$.  Thus the number of records
grows slowly with time; this result has the obvious consequence that records
become less likely to occur at later times.

\section{MONTE CARLO SIMULATIONS}
\label{simulations}

To verify our theoretical derivations, Monte Carlo simulations were performed
for both the exponential and Gaussian temperature distributions.  Our
simulations typically involve $10^5$ realizations (days) over a minimum of
1000 years of observations, and continue until six record temperatures have
been achieved.  We use ``years'' consisting of $10^5$ days so that we
generate a sufficient number of record temperatures to have reasonable
statistics.  For our initial simulations, we used a stationary mean and
variance of $18^\circ$C and $5^\circ$C respectively, which are typical values
for the distribution of maximum daily temperatures in the spring or fall in
Philadelphia.  However, the numerical validation of our theoretical
distributions does not depend on the particular values of mean and variance.

The simulation errors using an exponential distribution for the $k^{\rm th}$
record (with $k=0,1,\ldots,5$) are: less than $3\times10^{-5}$ for
$\mathcal{P}_k(T)$ (Eq.~\eqref{prob-T-exp}) using a distribution with 100
bins; $8.3\times10^{-5}$ for $Q_n(0)$ (Eq.~\eqref{Qn0-exact});
$2.2\times10^{-3}$ (relative error) for the mean temperature of the $k^{\rm
  th}$ record temperature (Eq.~\eqref{Tk}); and 0.01 (relative error) for the
variance.  The Gaussian distribution yields fewer exact expressions for
comparison, but includes a relative error of $6.4\times10^{-3}$ for the mean
temperature of the $k^{\rm th}$ record temperature (Eq.~\eqref{Tk-asymp}),
$k=0\ldots5$.  For both the exponential and Gaussian distributions, the
probability of breaking a record temperature with time is well fit by the
form $1/(t+1)$, with an error of less than $9.2\times10^{-5}$.  These errors
decrease as the number of realizations increases, and the small errors for
simulations with $10^5$ realizations confirm the correctness of the
theoretical distributions.

Monte Carlo simulations were also performed to explore the effect of temporal
correlations in daily temperatures on the frequency statistics of
record-temperature events and the magnitude of successive record
temperatures.  This topic will be discussed in detail in Sec.~\ref{corr}.  We
used the Fourier filtering analysis method \cite{ben-Avraham_Havlin00bk,MSSS}
to generate power-law correlations between daily temperature data for years
consisting of $10^4$ days over 200 years and for several values of the
exponent in the power law of the temporal correlation function.

\section{RECORD TEMPERATURE DATA}
\label{TRD}

Between 1874 and 1999, a total of 1707 record highs (4.68 for each day on
average) and 1343 record lows (3.68 for each day) occurred in Philadelphia
\cite{future}.  Because the temperature was reported as an integer, a
temperature equaling a current record could represent a new record if the
measurement was more accurate.  With the less stringent definition that a new
record either exceeds or {\em equals} the current record, the number of
record high and record low events over 126 years increased from 1707 to 2126
and from 1343 to 1793, respectively.  However, this alternative definition
does not qualitatively change the statistical properties of record
temperature events.

\begin{figure}[ht]
  \vspace*{0.cm}
  \includegraphics*[width=0.45\textwidth]{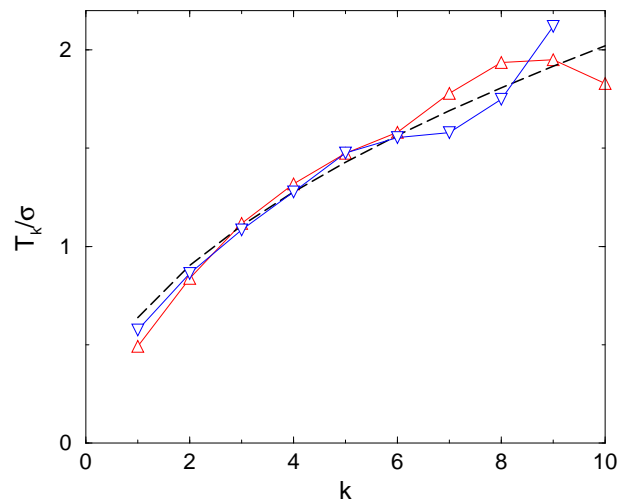} 
  \caption{(Color online) Average $k^{\rm th}$ record high ($\triangle$) and record low 
    ($\nabla$) temperature for each day, divided by the daily temperature
    dispersion, versus $k$ (from the Philadelphia temperature data).  The
    dashed curve is $T_k/\sigma=1.15\sqrt{k}$.
    \label{Tk-vs-k}}
\end{figure}

\begin{figure}[!ht]
  \vspace*{0.cm}
  \includegraphics*[width=0.45\textwidth]{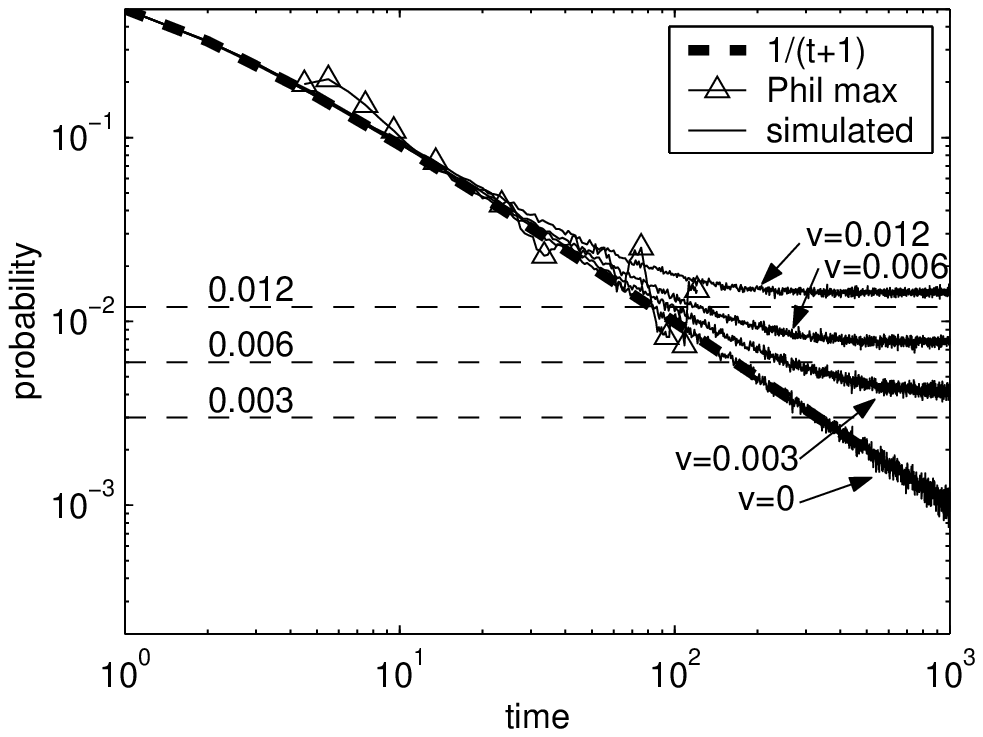} 
  \includegraphics*[width=0.45\textwidth]{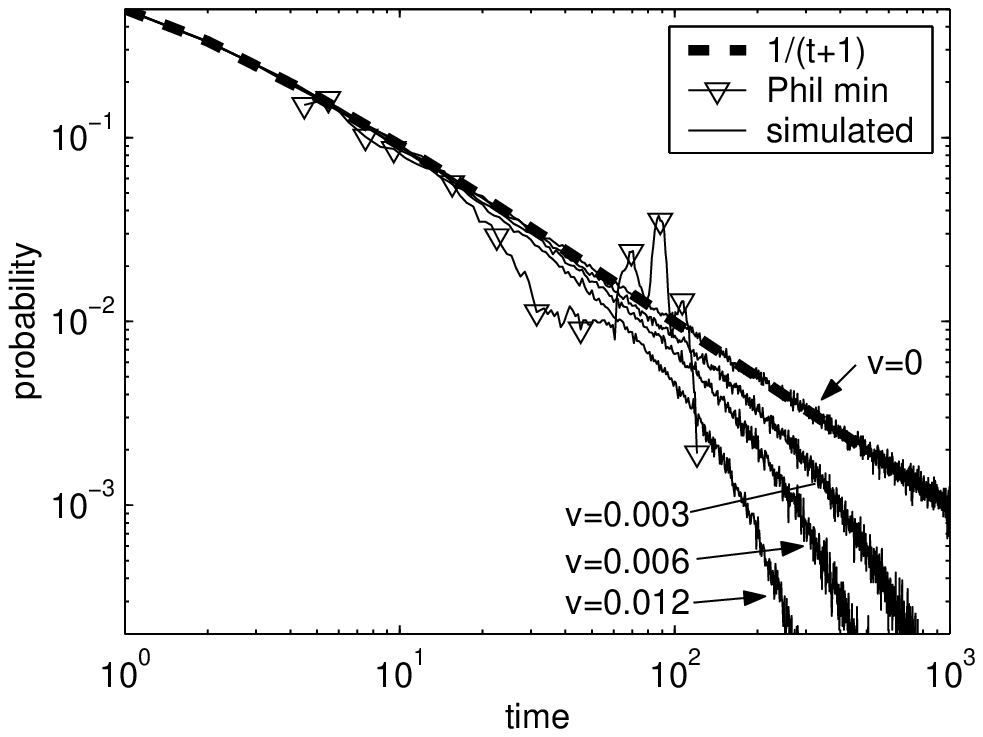} 
  \caption{Probability that a record high temperature (top) or record low
    (bottom) occurs at a time $t$ (in years) after the start of observations.  The
    symbols $\triangle$ and $\nabla$ are 10-point averages of Philadelphia
    data from 1874 to 1999 for ease of visualization.  Simulated data were
    produced by a stationary Gaussian distribution ($v=0$), or where the mean
    increases according to $v=0.003$, 0.006, or 0.012C year$^{-1}$.  
    The stationary data fit the theoretical expectation of $1/(t+1)$ (thick dashed line), while
    warming leads the distribution to asymptote to a constant probability 
    (thin dashed lines).
    \label{max-warming}}
\end{figure}

To compare with our theory, first consider the size of successive record
temperatures.  According to Eq.~(\ref{Tk-G}), the $k^{\rm th}$ record high
(and record low) temperature should be proportional to $\sqrt{2k\sigma^2}$.
Because the mean temperature for each day has already been subtracted off,
here $T_k$ denotes the absolute value of the difference between the $k^{\rm
  th}$ record temperature and the zeroth record.  To have a statistically
meaningful quantity, we compute $T_k/\sigma_\alpha$ for each day of the year,
and then average over the entire year; here the subscript $\alpha=h,l$
denotes the daily dispersion for the high and low temperatures, respectively.
As shown in Fig.~\ref{Tk-vs-k}, the annual average for $T_k/\sigma_\alpha$ is
consistent with $\sqrt{k}$ growth for both the record high and record low
temperature.  Up to the $6^{\rm th}$ record, both data sets are quite close,
and where the data begin to diverge, the number of days with more than 6
records is small---69 for high temperatures and 26 for low temperatures.

\begin{figure}[!ht]
  \vspace*{0.cm}
  \includegraphics*[width=0.45\textwidth]{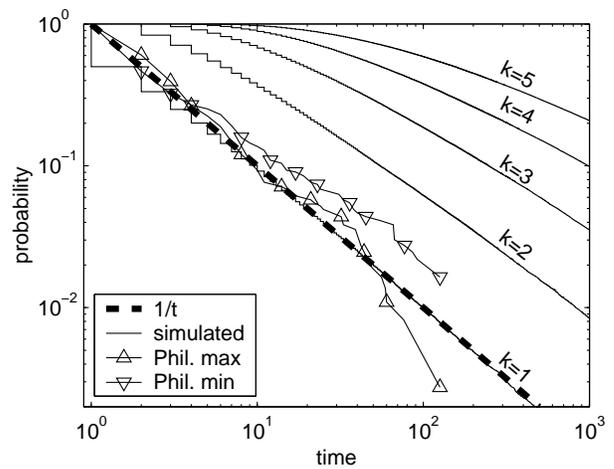}
  \caption{Probability that the $k^{\rm th}$ record high temperature occurs 
    at time $t$  (in years) or later, using simulated data (solid curves).  The $k=1$
    simulated data closely match the asymptotic theoretical distribution of
    $1/t$ (dashed line).  Also shown are the $k=1$ data for record high
    temperatures ($\Delta$) and record low temperatures ($\nabla$) for the
    Philadelphia data.
    \label{Q-warming}}
\end{figure}

Finally, we study the evolution of the frequency of record temperature days
as a function of time.  As discussed in Sec.~\ref{ETR}, the number of records
in the $t^{\rm th}$ year of observation (since 1874) should be $365/t$.  In
spite of the year-to-year fluctuations in the number of records, the
prediction $365/t$ fits the overall trend (Fig.~\ref{max-warming}).  We also
examine the distribution of waiting times between records.  Since the amount
of data is small, it is useful to study the cumulative distribution,
$\mathcal{Q}_n(k)\equiv \sum_{m=n}^\infty Q_m(k)$, defined as the probability
that the time between the $k^{\rm th}$ and the $(k+1)^{\rm st}$ record
temperatures on a given day is $n$ years or larger.  As shown in
Fig.~\ref{Q-warming}, the agreement between the Philadelphia data and the
theoretical prediction from Eq.~(\ref{Qn0-G-asymp}), $\mathcal{Q}_n(0)\propto
1/n$ is quite good.  The Monte Carlo simulations match the theoretical
prediction nearly exactly, with an rms error of $9\times10^{-5}$.

In summary, the data for the magnitude of temperature jumps at each
successive record, the frequency of record events, and the distribution of
times between records are consistent with the theoretical predictions that
arise from a Gaussian daily temperature distribution with a stationary mean
temperature.

\section{SYSTEMATICALLY CHANGING TEMPERATURE}
\label{SCT}

We now study how a systematically changing average temperature affects the
evolution of record temperature events.  For global warming, we assume that
the mean temperature has a slow superimposed time dependence $vt$, with $v>0$
and where $t$ is the time (in years) after the initial observational year.

\subsection{Exponential distribution}
Again, as a warm-up exercise, we first consider the idealized case of an
exponential daily temperature distribution,
\begin{equation}
\label{p-inc}
p(T;t)= 
\begin{cases} e^{-(T-vt)} & T>vt  \\
0 & T<vt\,,
\end{cases}
\end{equation}
where we set the characteristic temperature scale $\mathcal{T}$ to 1 for
simplicity.  In these units, both $T$ and $vt$ are dimensionless.  With this
distribution, the recursion Eq.~(\ref{Tk-gen}) for successive record
temperatures becomes
\begin{equation}
\label{Tk-gen-v}
 T_{k+1} \equiv \frac{\int_{T_k}^\infty  T\, e^{-(T-vt_{k\!+\!1})}\, dT}
{\int_{T_k}^\infty  e^{-(T-vt_{k\!+\!1})}\, dT}\,.
\end{equation}
The factor $e^{vt_{k+1}}$ appears in both the numerator and denominator and
thus cancels.  As a result, $T_k=k+1$, {\em independent\/} of $v$.  Thus a
systematic temperature variation---either global warming or global
cooling---does not affect the magnitude of the jumps in successive record
high temperatures.  This fact was verified by numerical simulations with an
exponential distribution, where the distributions of $\mathcal{P}_k(T)$ for
$v=0.012^\circ$C year$^{-1}$ and $v=0$ match to within a few percent for
$k=0\ldots5$.

On the other hand, a systematic temperature dependence does affect the time
between records.  Suppose that the current record high temperature of $T_k$
was set in year $t_k$.  Then the exceedance probability at time $t_k+j$ is
\begin{eqnarray}
\label{pg-exp-move}
p_>(T_k;t_k+j)&=& \int_{T_k}^\infty e^{-[T-v(t_k+j)]}\,dT\nonumber \\
&=&e^{-(T_k-vt_k)}\,e^{jv}\equiv X\,e^{jv}\,.
\end{eqnarray}
The exceedance probability is thus either enhanced or suppressed by a factor
$e^v$ due to global warming or cooling, respectively, for each elapsed year.
The probability $q_n(T_k)$ that a new record high temperature occurs $n$
years after the previous record $T_k$ at time $t_k$ is
\begin{equation}
\label{prod-inc}
q_n(T_k)=\, e^{nv}\, X\,\prod_{j=1}^{n-1} (1-e^{jv}\,X),
\end{equation}
with $q_1(T_k)= e^v\,X$; this generalizes Eq.~(\ref{qn}) to incorporate a
global climatic change.

For the case of global warming ($v>0$), each successive term in the product
decreases in magnitude and there is a value of $j$ for which the factor
$(1-e^{jv}\, X)$ is no longer positive.  At this point, the next temperature
must be a new record.  Thus we (over)estimate the time until the next record
after $T_k$ by the criterion $(1-e^{jv}\, X)=0$, or $j= (T_k-vt_k)/v\sim
(k/v)-t_k$.  Since this value of $j$ also coincides with $t_{k+1}-t_k$ by
construction, we obtain $t_k\sim k/v$.  Thus the time between consecutive
records asymptotically varies as $t_{k+1}-t_k\sim 1/v$.  This conclusion
agrees with a previous mathematical proof of the constancy of the rate of new
records when a linear temporal trend is superimposed on a set of continuous
iid variables \cite{trend}; a different approach to deal with a linear trend
is given in \cite{B}.

If global warming is slow, the waiting time between records will initially
increase exponentially with $k$, as in the case of a stationary temperature,
but then there will be a crossover to the asymptotic regime where the waiting
time is constant.  We estimate the crossover time by equating the two forms
for the waiting times, $t_{k+1}-t_k=e^{(k+1)}$ (stationary temperature) and
$t_{k+1}-t_k=1/v$ (increasing temperature), to give $k*\approx - \ln v$.  Now
the average annual high temperature in Philadelphia has increased by
approximately $1.94^\circ$C over 126 years.  The resulting warming rate of
$0.0154^\circ$C per year then gives $k*\approx 3.6$.  Thus the statistics of
the first 3.6 record high temperatures should be indistinguishable from those
in a stationary climate, after which record temperatures should occur at a
constant rate.  Since the average number of record high temperatures for a
given day is 4.7 and the time until the next record high is very roughly
$e^{5.7}-e^{4.7}\approx 190$ years, we are still far from the point where
global warming could have an unambiguous effect on the frequency of record
high temperatures.

For global cooling ($v<0$), the waiting time probability becomes
\begin{equation}
\label{prod-dec}
q_n(T_k)=\prod_{j=1}^{n-1} (1-e^{-jw}\,Y)\, e^{-nw}\, Y\,,
\end{equation}
with $q_1(T_k)= e^{-w}\,Y$, where $w\equiv|v|$ is positive, and
$Y=e^{-T_k-wt_k}$.  We estimate the above product by the following simple
approach.  When $jw<1$, then $e^{-jw}\ll 1$, and each factor within the
product is approximately $(1-Y)$.  Consequently, for $nw>1$, each term in the
product approximately equals $(1-Y)$ for $j<n*=1/w$, while for $j>n*$,
$e^{-jw}\approx 0$, and the later terms in the product are all equal to 1.
Thus
\begin{equation}
\label{q-inc}
q_n(T_k)\sim
\begin{cases} (1-Y)^n\, e^{-nw}\,Y & n<n*  \\
(1-Y)^{1/w}\,e^{-nw} \,Y & n>n*\,.
\end{cases}
\end{equation}

Using this form for $q_n$, we find, after straightforward but slightly
tedious algebra, that the dominant contribution to the waiting time until the
next record temperature, $t_{k+1}-t_k =\sum_{n=1}^\infty n q_n$, comes from
the terms with $n<n*$ in the sum.  For the case slow global cooling, we
thereby find
\begin{eqnarray}
\label{tk-cool}
t_{k+1}-t_k &\sim& \frac{1/Y}{[1+w(1/Y-1)]^2}
\sim 1/Y \nonumber \\ &=& e^{T_k+wt_k}\,.
\end{eqnarray}
Since $t_{k+1}-t_k\approx dt/dk$ and using $T_k\sim k$, Eq.~(\ref{tk-cool})
can be integrated to give $(1-e^{-wt_k})=w(e^k-1)$.  As long as the
right-hand side is less than 1, a solution for $t_k$ exists.  In the converse
case, there is no solution and thus no additional record highs under global
cooling, or equivalently, no more record lows for global warming.  For small
$w$ and in the pre-crossover regime where $e^k\approx t_k$, the criterion for
no more records reduces to $t>1/w$.  If the daily low temperature in
Philadelphia also experienced a warming rate of $0.0154^\circ$C per year,
then there should be no additional record low temperatures after about 36
years of observations.  However, the daily low temperatures do not show a
long-term systematic variation, so new record lows should continue to occur,
as is observed.

\subsection{Gaussian distribution}
We now treat the more realistic case where a systematic temperature variation
is superimposed on a Gaussian daily temperature distribution, as embodied by
\begin{equation}
p(T;t)=\frac{1}{\sqrt{2\pi \sigma^2}}\, e^{-(T-vt)^2/2\sigma^2}\,.
\end{equation}
The details of the effects of a systematic temperature variation on the
statistics of record temperatures are tedious and we merely summarize the
main results.  We assume a slow systematic variation, $T_k-vt\gg 0$, so that
an asymptotic analysis will be valid.  Under this approximation, both global
warming or global cooling lead to the following recursion for $T_k$, to
leading order:
\begin{equation}
\label{Tk-warm-G}
T_{k+1}-T_k \sim \frac{\sigma^2}{T_k}\left(1+\frac{vt}{T_k}\right)\,.
\end{equation}
The term proportional to $vt$ in Eq.~(\ref{Tk-warm-G}) is subdominant, so
that $T_k$ still scales as $\sim\sqrt{2k\sigma^2}$, for both global warming
and global cooling.

Next we determine the times between successive record high temperatures.  The
basic quantity that underlies these waiting times is again the exceedance
probability, when the current record is $T_k$ and the current time is
$t_k+j$.  Following Eq.~(\ref{pg-G}), this exceedance probability is
\begin{equation}
p_>(T_k;t_k+j)\sim \frac{1}{2}\,\,{\rm erfc}\left(\frac{T_k-v(t_k+j)}{\sqrt{2\sigma^2}}\right)\,.
\end{equation}
In the asymptotic limit where the argument of the complementary error
function is large, the controlling factor in $p_>$ is 
\begin{equation}
e^{-[T-v(T_k+j)]^2/2\sigma^2}\approx e^{-(T-vt_k)^2/2\sigma^2}\, e^{vj(T-vt_k)/\sigma^2}\,.
\end{equation}

The crucial point is that the latter form for the exceedance probability has
the same $j$ dependence as in the exponential distribution
(Eq.~(\ref{pg-exp-move})).  Thus our arguments for the role of global warming
with an exponential daily temperature distribution continue to apply.  In
particular, the time between successive records initially grows as
$\sqrt{4\pi k}\, e^k$, but then asymptotically approaches the constant value
$1/v$.  As a result, the time before global warming measurably influences the
frequency of record high and record low temperatures will be similar for both
the exponential and Gaussian temperature distributions.

Monte Carlo simulations were performed for warming rates $v=0.003$, $0.006$,
and $0.012^\circ$C/year, where the middle case corresponds to the accepted
rate of global mean warming of 0.6$^\circ$C for the $20^{\rm th}$ century
\cite{IPCC01bk}.  Unlike the exponential distribution simulations, for the
Gaussian distribution $\mathcal{P}_k(T)$ is slightly different in the cases
of no warming and warming (Fig.~\ref{Pk}).

Figure~\ref{max-warming} shows the results of numerical simulations using the
Gaussian distribution with $10^5$ realizations for the three warming rates.
For the stationary case ($v=0$), the probability of breaking a record after
$t$ years closely follows the theoretical expectation of $1/(t+1)$.  For
warming, the rate of breaking a record high (Fig.~\ref{max-warming}, top)
ultimately asymptotes to a constant frequency of approximately $1.25v$ by
$10^4$ years.  Given our crude calculation following Eq.~(\ref{prod-inc})
that the time between records is $1/v$, the agreement between the observed
rate of $1.25v$ and our estimate of $v$ is gratifying.  As also predicted in
our theory, the probability of breaking a record low temperature under global
warming precipitously decays after a few hundred years
(Fig.~\ref{max-warming}, bottom); eventually record low temperatures simply
stop occurring in a warming world.

\section{ROLE OF TEMPORAL CORRELATIONS}
\label{corr}

Thus far our presentation has been based on {\em independent\/} daily
temperatures---no correlations between temperatures on successive days.
However, from common experience we know that local weather consists of
multi-day patterns within which smaller temperature variations occur.
Anecdotally, the temperature tomorrow will be close to the temperature today.
In fact, it has been found in global climatological data that correlations
between temperatures on two widely separated days decay as a power law in
the separation \cite{Bunde_ea05prl}.  Here we quantify these correlations for
the Philadelphia data and then discuss the potential ramifications of these
correlations on the frequency of record temperature events.

\subsection{Daily temperature correlation data}

From the Philadelphia data, we compute the normalized interday temperature
correlation function defined as
\begin{eqnarray}
\label{corr-def}
c_\alpha(i,j)=\frac{\langle T_i T_j\rangle - \langle T_i \rangle\langle T_j\rangle}
{\langle T_i^2\rangle - \langle T_i \rangle^2}\,.
\end{eqnarray}
Here $i$ and $j>i$ denote the $i^{\rm th}$ and $j^{\rm th}$ days of the year,
$T_i$ is the temperature on the $i^{\rm th}$ day, and $\langle T_i\rangle$ is
its average value over the 126 years of data, while the index $\alpha=h,m,l$
denotes the high, middle, and low temperature for each day.  If $i$ is a day
near the end of the year, then $T_j$ will refer to a temperature in the
following year when the separation between the two days exceeds $(365-i)$.
According to Eq.~\eqref{corr-def}, if the temperatures $T_i$ and $T_j$ are
both greater than or both less than the respective average temperatures for
days $i$ and $j$, then there is a positive contribution to the correlation
function.  Thus $c_\alpha(i,j)$ measures systematic temperature deviations
from the mean on these two days.  For convenience, we normalize the
$c_\alpha$ so that they all equal 1 when $|i-j|=0$.

\begin{figure}[ht]
  \vspace*{0.cm}
  \includegraphics*[width=0.45\textwidth]{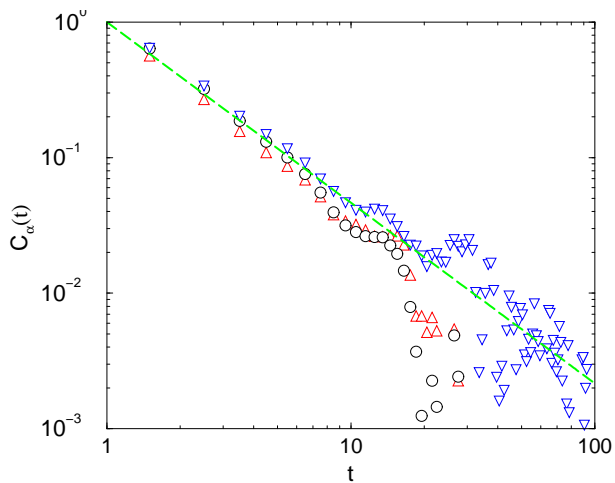} 
  \caption{(Color online) The correlation functions $C_\alpha(t)$ for high 
   ($\triangle$), middle ($\circ$), and low temperature ($\nabla$) versus
   time (in years).  The straight line of slope $-4/3$ is a guide for the eye.
   \label{correlation-norm}}
\end{figure}

The correlation functions depend primarily on the separation between the two
days, $|i-j|$, and weakly on the initial day $i$.  To obtain a succinct
measure of the temperature correlation over a year, we define the annual
average correlation function
\begin{eqnarray}
{C_\alpha(t)}\equiv \sum_{i=1}^{365}c_\alpha(i,i+t)\,.
\end{eqnarray}
All three correlations functions are consistent with a power law decay
$C_\alpha(t)\sim t^{-\gamma}$ (Fig.~\ref{correlation-norm}).  Over a range of
approximately 1--20 days, the best fit value of $\gamma$ is 1.29 for $C_h$
(which remains strictly positive until 36 days) and $\gamma =1.44$ for $C_m$
(which remains strictly positive until 41 days).  The correlation function
$C_l$ is visibly distinct and remains strictly positive until 149 days, with
a best-fit exponent of $\gamma=1.36$.  These power-law decays in the
temperature correlation functions are consistent with the previous results of
Ref.~\cite{Bunde_ea05prl}.  However, the exponent value that we observe,
approximately 4/3, is considerably larger than that reported in
Ref.~\cite{Bunde_ea05prl}.  The time integrals of the high-, middle-, and
low-temperature correlation functions are 1.78, 2.04 and 5.16 respectively.
We may therefore view 1.78 as the average length of an independent
high-temperature event, and correspondingly $365/1.78\approx 205$ as the
number of effective independent ``days'' for high temperatures.  Parallel
results hold for middle and low temperatures.  These numbers provide a
feeling for the extent of multiday weather patterns because of temperature
correlations.
\subsection{Simulations with correlated daily temperatures}

\begin{figure}[!ht]
  \vspace*{0.cm}
  \includegraphics*[width=0.45\textwidth]{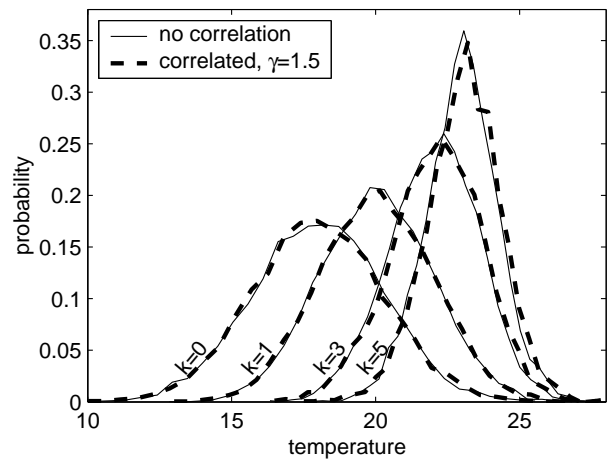}
  \caption{Simulation data for the probability distribution of the
    $k^{\rm th}$ record high temperature in degrees Celsius, $\mathcal{P}_k(T)$, where
    daily temperatures are uncorrelated (solid line) and power-law correlated
    with exponent $1.5$.
    (dashed line).
    \label{corr_temp}}
\end{figure}

\begin{figure}[!ht]
  \vspace*{0.cm}
  \includegraphics*[width=0.45\textwidth]{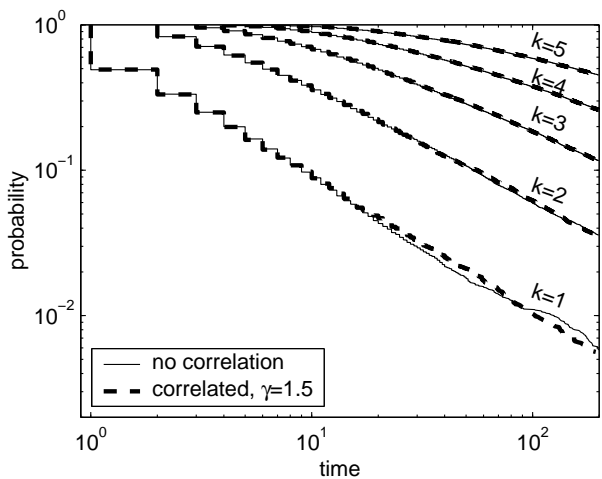}
  \caption{Probability that the $k^{\rm th}$ record high temperature occurs 
    at time $t$ (in years) or later, using uncorrelated (solid line)
    and power-law correlated daily temperatures (dashed line).
    \label{corr_time}}
\end{figure}

To determine if these correlations affect the frequency and magnitude of
record temperature events, we performed Monte Carlo simulations in which
daily temperatures had temporal correlations that matched the data discussed
above.  We generate such correlated data using the Fourier filtering method of
Refs.~\cite{ben-Avraham_Havlin00bk,MSSS} with a correlation function of the form
\begin{eqnarray}
{C_\alpha(t)} = t^{-\gamma}
\end{eqnarray}
for a range of $\gamma$ values around the observed value of $1.3$--$1.4$.
Due to the computational demands of generating correlated data, simulations
of years consisting of $10^4$ days for 200 years were performed, which are
less extensive than our simulations for uncorrelated temperatures.  We find
that the statistics of the time between record temperature events and the
magnitude of successive record temperatures are virtually identical to those
obtained when the temperature is an independent identically-distributed
random variable.  (Figs.~\ref{corr_temp} and \ref{corr_time}).  Our results
are also not sensitive to the value of the decay exponent $\gamma$ of the
correlation function, within our tested range of $\gamma \in [0.5,1.5]$.
We conclude that the discussion in Secs.~\ref{ETR}--\ref{SCT}, which
assumed uncorrelated day-to-day temperatures, can be applied to real
atmospheric observations, where daily temperatures are correlated.  It is
worth mentioning, however, that interday correlations do strongly affect the
statistics of successive extremes in temperatures \cite{corr-extreme}.

\subsection{Correlations between record temperature events}

While temperature correlations do not affect record statistics for a given
day, these correlations should cause records to occur as part of a heat wave
or a cold snap, rather than being singular one-day events.  As a matter of
curiosity, we studied the distribution of times (in days) between successive
record events, as well as the distribution of streaks (consecutive days) of
record temperatures from the time history of all record temperature events.

Because the number of record temperatures decreases from year to year, these
time and streak distributions are not stationary.  We compensate for this
non-stationarity by rescaling so that data for all years can be treated on
the same footing.  For example, for the distribution of times between
successive records, we rescale each interevent time by the average time
between records for that year.  Thus, for example, if two successive records
occurred 78 days apart in a year where 5 record temperature events occurred
(average separation of 73 days), the scaled separation between these two
events is $\tau=78/73\approx 1.068$.  For the length of record streaks, we
similarly rescaled each streak by the average streak length in that year,
assuming record temperature events were uncorrelated.

\begin{figure}[ht]
  \vspace*{0.cm} \includegraphics*[width=0.4\textwidth]{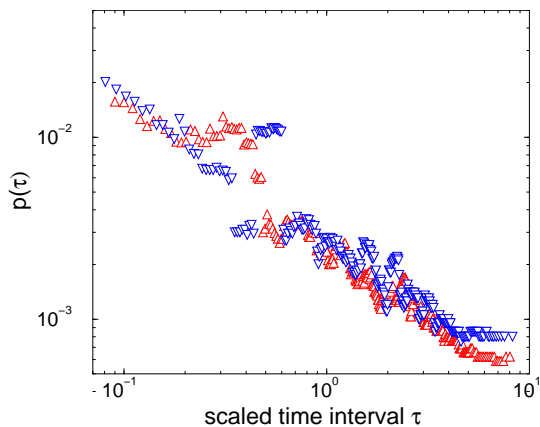}
  \caption{(Color online) Distribution of times $p(\tau)$ between successive record temperature
    events ($\triangle$ record highs, $\nabla$ record lows).  The times are 
    scaled by the average time between record events for each year.  
    \label{record-time-dist}}
\end{figure}

The distribution of times between successive record temperature days decays
slower than exponentially (Fig.~\ref{record-time-dist}); the latter form
would occur if record temperature events were uncorrelated.  In a similar
vein, we observe an enhanced probability for records to occur in streaks.
Since record streaks are rare, we can only make the qualitative statement
that the streak distribution is different than that from uncorrelated data.
Our basic conclusion is that interday temperature correlations do affect
statistical features of successive record temperature events but do not
affect the statistics of record temperatures on a given day, where events are
more than one year apart.

\section{DISCUSSION}
\label{disc}

Two basic aspects of record temperature events are the size of the
temperature jump when a new record occurs and the separation in years between
successive records on a given day.  We computed the distribution functions
for these two properties by extreme statistics reasoning.  For the Gaussian
daily temperature distribution, we found that (i) the $k^{\rm th}$ record high
temperature asymptotically grows as $\sqrt{k}\,\sigma$, where $\sigma$ is the
dispersion in the daily temperature, and (ii) record events become
progressively less likely, with the typical time between the $k^{\rm th}$ and
$(k+1)^{\rm st}$ record growing as $\sqrt{k}\,e^k$.  This latter result is
independent of $\sigma$ so that systematic changes in temperature variability
should not affect the time between temperature records.

{}From these predictions, the distribution of waiting times between two
successive records on a given day has an inverse-square power-law tail, with
a divergent average waiting time.  Furthermore, the number of record events
in the $t^{\rm th}$ year of observations decays as $t^{-1}$
\cite{ABN,Glick78amm,Sibani,KJ,Benestad03cr}.  These theoretical predictions
agree with numerical simulations and with data from 126 years of observations
in Philadelphia.  Another important feature is that the annual frequency of
record temperature events is not measurably influenced by interday power-law
temperature However, these correlations do play a significant role at shorter
time scales.

Our primary result is that we cannot {\em yet\/} distinguish between the
effects of random fluctuations and long-term systematic trends on the
frequency of record-breaking temperatures with 126 years of data.  For
example, in the $100^{\rm th}$ year of observation, there should be
$365/100=3.65$ record-high temperature events in a stationary climate, while
our simulations give $4.74$ such events in a climate that is warming at a
rate of $0.6^\circ$C per 100 years.  However, the variation from year to year
in the frequency of record events after 100 years is larger than the
difference of $4.74-3.65$, which should be expected because of global warming
(Fig.~\ref{max-warming}).  After 200 years, this random variation in the
frequency of record events is still larger than the effect of global warming.
On the other hand, global warming already does affect the frequency of
extreme temperature events that are defined by exceeding a fixed threshold
\cite{Yan_ea02cc,Mearns_ea84jcam,Hansen_ea88jgr,Katz_Brown92cc,
  Columbo_ea99jc,Unkasevic_ea05tac}.

\begin{figure}[ht]
  \vspace*{0.cm} \includegraphics*[width=0.4\textwidth]{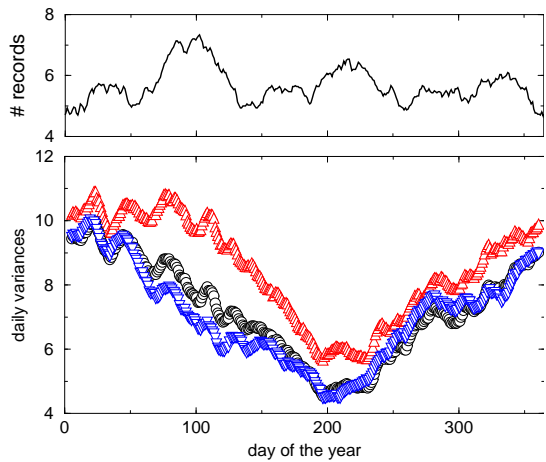}
  \caption{(Color online) Number of high-temperature records for each day of
    the year, averaged over a 30-day range (top).  Below are the variances
    in the high ($\triangle$), middle ($\circ$), and low ($\nabla$)
    temperatures for each day averaged over a 10-day range.
    \label{daily-var}}
\end{figure}

While the agreement between our theory and the data for record temperature
statistics is satisfying, there are various facts that we have either glossed
over or ignored.  These include: (i) a significant difference between the
number of record high and record low events---1705 record high events and
only 1346 record low events have occurred the 126 years of data.  (ii) A
propensity for record high temperatures in the early spring.  This
seasonality is illustrated both by the number of records for each day of the
year and by the daily temperature variance $\sigma_i \equiv \sqrt{\langle
  T^2_i\rangle - \langle T_i\rangle^2}$, where $\langle T_i\rangle$ and
$\langle T^2_i\rangle$ are the mean and mean-square temperatures for the
$i^{\rm th}$ day (Fig.~\ref{daily-var}).  (iii) The potential role of a
systematically increasing variability on the frequency of records.  For the
last point, Krug \cite{krug} has shown that for an exponential daily
temperature distribution whose width is increasing linearly with time, the
number of record events after $t$ years grows as $(\ln t)^2$, intermediate to
the $\ln t$ growth of a stationary distribution and linear growth when the
average temperature systematically increases.  (iv) Day/night or high/low
asymmetry \cite{asymmetry}.  That is, as a function of time there are more
days whose highs exceeds a given threshold and fewer days whose high is less
than a threshold.  Paradoxically, however, there are fewer days whose lows
exceed a given temperature and more days whose lows are less than a given
temperature.  Since highs generally occur in daytime and lows in nighttime,
these results can be restated as follows: the number of hot days is
increasing {\it and\/} the number of cold nights is increasing.  We don't
know how this latter statement fits with the phenomenon of global warming.

Another caveat is that our theory applies in the asymptotic limit, where each
day has experienced a large number of record temperatures over the
observational history.  The fact that there are no more than 10 record events
on any single day means that we are far from the regime where the asymptotic
limit truly applies.  Finally, and very importantly, it would be useful to
obtain long-term temperature data from many stations to provide a more
definitive test of our predictions.

\acknowledgements{ We thank D. ben-Avraham for insightful discussions, C.
  Forest for helpful advice, and P. Huybers and R. Katz for constructive
  manuscript suggestions and literature advice.  We also thank J. Krug for
  informing us of Refs.~\cite{trend,B} about the effect of linear trends on
  record statistics after this work was completed, S. Majumdar for helpful
  discussions that led to the derivations given in
  Eqs.~\eqref{indi}--\eqref{PRn}, Z. Racz for making us aware of literature
  on high/low asymmetry, and a referee for constructive suggestions.
  Finally, we gratefully acknowledge financial support from DOE grant
  W-7405-ENG-36 (at LANL) and NSF grants DMR0227670 and DMR0535503 (at BU).}

\end{document}